\documentclass[twocolumn,prd,aps,epsfig]{revtex4}
\usepackage{graphicx}
\begin{document}
\renewcommand{\baselinestretch}{1.2}

\def\del{\partial }

\null\noindent{TUHEP-TH-03142}
\title{Distinguishing technicolor models via $t\bar{t}$ productions
at polarized photon colliders}
\bigskip
\author{Bin Zhang,~~Yuanning Gao}
\address{Center for High Energy Physcs, 
Tsinghua University, Beijing 100084, China}
\author{Yu-Ping Kuang}
\address{ CCAST (World Laboratory), P.O. Box 8730, Beijing 100080, China}
\address{Center for High Energy Physics,
Tsinghua University, Beijing 100084, China\footnote{Mailing address}}

\bigskip\bigskip

\begin{abstract}

We study top quark pair productions at a polarized photon collider
from an $e^+e^-$ linear collider (LC) in various improved
technicolor model, namely, the one-family walking technicolor
model, the top-color-assisted technicolor model, and the
top-color-assisted multiscale technicolor model. Recent constraint
on the top-pion mass from the precision data of $R_b$ is
considered. It is shown that, considering only the statistical errors,
a polarized photon collider from a 500 GeV LC with an integrated
luminosity of 500 fb$^{-1}$ is sufficient for distinguishing the
three improved technicolor models experimentally.

\vspace{0.2cm}
\null\noindent PACS number(s): 12.60.Nz, 13.40.-f,
14.65.Ha
\end{abstract}
\maketitle

\bigskip\bigskip

Although the standard model (SM) has successfully passed the
precision electroweak tests, its elecrtoweak symmetry breaking
(EWSB) mechanism is still unclear. The Higgs boson has not been
found, and the LEP2 bound on the Higgs boson mass is 114.4 GeV
\cite{PDG}. Furthermore, the SM Higgs sector suffers from the
well-known problems of {\it triviality} and {\it unnaturelness}
arising from the elementary Higgs field. There have been many new
physics models of the EWSB mechanism proposed for avoiding the
above problems. An attractive idea of completely avoiding {\it
triviality} and {\it unnatureness} is to abandon the elementary
Higgs field(s), such as various kinds of improved technicolor
models \cite{AT,TOPCTC,TOPCMTC}, top quark seesaw models
\cite{seesaw}, and certain little Higgs models \cite{littleHiggs}. In a
previous paper \cite{ZKYWL}, we studied the tests of various
improved technicolor models via top quark pair productions at high
energy photon colliders. Since the pseudo Goldstone bosons (PGBs)
coupling to the top quark in different improved technicolor models
are quite different, we showed that different improved technicolor
models can be distinguished experimentally through top quark pair
productions at an unpolarized photon collider built from a
$\sqrt{s}=1.5$ TeV $e^+e^-$ linear collider (LC) \cite{ZKYWL}. In
this paper, we study the same processes at polarized photon
colliders, and take into account the recent bound on the top-pion
mass from the precision data of $R_b$ \cite{YKWL}. We shall see
that, considering only statistical uncertainties, a polarized photon
collider built from a $\sqrt{s}=500$ GeV LC with an integrated
luminosity of 500 fb$^{-1}$ is sufficient for distinguishing the improved
technicolor models.

It has been shown that the polarization of the initial laser beam
and the electron beam will significantly affect the photon
spectrum at the photon collider \cite{Ginzburg}. Let $P_c$ be the
polarization of the initial laser, $\lambda_e$ be the polarization
of the electron in the first beam, and $\tilde{P}_c$ and
$\tilde{\lambda}_e$ be the corresponding parameters in the second
beam, respectively. It is shown in Ref.\cite{Ginzburg} that the
colliding photons will peak in a narrow region near the high
energy end ($80\%$ of the electron energy) if $2\lambda_e P_c=-1$.
This improves the monochromatization and enhances the effective
energy of the colliding photons at the photon collider. We shall
see that this effect leads to the possibility of distinguishing
different iomproved technicolor models at a polarized photon
collider built from a $\sqrt{s}=500$ GeV LC.

Let $E_e,~\omega_0,~\omega$,~and $\sqrt{s}(\sqrt{\hat s})$ be the
incident elecrton energy, the laser-photon energy, the
backscattered photon energy, and the center-of-mass energies of
$e^+e^-$($\gamma\gamma$), respectively. The photon luminosity
$dL_{\gamma\gamma}$ is \cite{luminosity}
\begin{eqnarray}                            
dL_{\gamma\gamma}=2zdz\int^{x_{max}}_{z^2/x_{max}}\frac{dx}{x}
F_{\gamma/e}(x)F_{\gamma/e}(z^2/x).
\end{eqnarray}
with $z=\sqrt{\hat{s}/s}$, and $x_{max}=\omega_{max}/E_e$,
\begin{eqnarray}
F_{\gamma/e}(x)&=&\frac{1}{D(\xi)}\bigg[1-x+\frac{1}{1-x}-
\frac{4x}{\xi(1-x)}+\frac{4x^2}{\xi^2(1-x)^2}\nonumber\\
&&-2\lambda_eP_c\bigg(\frac{x}{1-x}-\frac{2x^2}{(1-x)^2\xi}\bigg)(2-x)\bigg],\nonumber
\end{eqnarray}
and
\begin{eqnarray}
D(\xi)=\bigg(1-\frac{4}{\xi}-\frac{8}{\xi^2}\bigg)\ln(1+\xi)
+\frac{1}{2}+\frac{8}{\xi}-\frac{1}{1(1+\xi)^2},\nonumber
\end{eqnarray}
where $\xi=4E_e\omega_0/m_e^2$. In order to avoid the creation of
$e^+e^-$ pairs by the interaction of the incident and
backscattered photons, $\xi$ should not be larger than 4.8
\cite{Ginzburg}. As in Ref.\cite{ZKYWL}, we take $\xi=4.8$, then
$x_{max}\approx 0.83$ and $D(\xi)=1.8$. The formula for the
counting rate of $\gamma\gamma\to X$ at a polarized photon
collider has been given in Ref.\cite{Ginzburg}. It is
\begin{eqnarray}                           
d\dot{N}_{\gamma\gamma\to X}&=& dL_{\gamma\gamma}( d\sigma+\Lambda
d\tau+\zeta_2\tilde{\zeta}_2
d\tau^a+\zeta_2 d\sigma_{20}\nonumber\\
&&+\tilde{\zeta}_2 d\sigma_{02}), \label{equNum1}
\end{eqnarray} where
$\zeta_i~(\tilde{\zeta}_i),~i=1,2,3$ are Stokes parameters
\cite{Ginzburg},
\begin{eqnarray}
&&\zeta_2=\frac{C_{20}}{C_{00}},\nonumber\\
&&C_{00}=\frac{1}{1-x}+1-x-4r(1-r)\nonumber\\
&&
\hspace{1cm}-2\lambda_e P_c r \xi(2r-1)(2-x)\nonumber
\\
&&C_{20}=2\lambda_e r
\xi[1+(1-x)(2r-1)^2]\nonumber\\
&&
\hspace{1cm}-P_c(2r-1)(\frac{1}{1-x}+1-x),\nonumber\\
&&x\equiv\frac{\omega}{E_e},~~~~~~r\equiv\frac{x}{\xi(1-x)},\nonumber
\end{eqnarray}
$\sigma_{ij},~i,j=0,1,2,3~$ are cross sections defined in
Ref.\cite{BGMS}, $\Lambda=\langle
\zeta_3\tilde\zeta_3-\zeta_1\tilde\zeta_1\rangle$ \cite{Ginzburg},
and $d\sigma$, $d\tau^a$ are \cite{Ginzburg}
\begin{eqnarray}                        
d\sigma&=&\frac{1}{4}(|M_{++}|^2+|M_{--}|^2+|M_{+-}|^2\nonumber\\
&&+|M_{-+}|^2)d\Gamma,\nonumber\\
d\tau^a&=&\frac{1}{4}(|M_{++}|^2+|M_{--}|^2-|M_{+-}|^2\nonumber\\
&&-|M_{-+}|^2)d\Gamma,
\label{amplitude}
\end{eqnarray}
in which the subscripts $+$ and
$-$ indicate that the photon helicity is $+1$ and $-1$,
respectively, and
\begin{eqnarray}
d\Gamma=\frac{(2\pi)^4}{4k\tilde{k}}\delta(k+\tilde{k}-\sum_f
p_f)\prod_f\frac{d^3p_f}{(2\pi)^3 2\epsilon_f},\nonumber
\end{eqnarray}
where $k,~\tilde{k}$ and $p_f$ are four momenta of the colliding
photons and the $f$-th final state particle, respectively. After
averaging over the azimuthal angles, $d\sigma_{20}$ and
$d\sigma_{02}$ vanish, and $\Lambda d\tau^a$ is negligibly small.
So Eq. (\ref{equNum1}) becomes
\begin{equation}                   
\label{equNum2} d\dot{N}_{\gamma\gamma\rightarrow
X}=dL_{\gamma\gamma}( d\sigma+\zeta_2\tilde{\zeta}_2 d\tau^a).
\end{equation}
The corresponding cross section $\sigma(s)$ at an LC with
center-of-mass energy $\sqrt{s}$ can be obtained by further
integrating Eq. (\ref{equNum2}) over the parameter $z$
\cite{luminosity}
\begin{eqnarray}                  
\sigma(s)=\int^{x_{max}}_{2m_t/\sqrt{s}}
\frac{d\dot{N}_{\gamma\gamma\to X}}{dz} dz. \label{sigma}
\end{eqnarray}

For the detection of the final state $t\bar{t}$, we know that
the dominant decay mode of the top quark is $t\to W^+b$. The $W$ boson
will then decay into either two leptons $l^+\nu_l$ or two quarks $q\bar{q}'$. We take the
hadronic mode $q\bar{q}'$ for detecting the final state signal. The branching ratio
of $W\to q\bar{q}'$ is $B(W\to q\bar{q}')=68\%$ \cite{PDG}. So the signal contains six jets
including two b-quark jets. To separate the six jets, we follow Ref. \cite{JLC}
to impose a cut on the clustering of jets. Let ${\hat y}$ be the jet-invariant-masses
normalized by the visible energy. The imposed cut is \cite{JLC}
\begin{eqnarray}                
{\hat y}>{\hat y}_{cut}=5\times 10^{-3}.
\label{ycut}
\end{eqnarray}

A possible background is $\gamma\gamma\to W^+W^-Z$ with $Z\to b\bar{b}$.
We know that, at $\sqrt{s}=500$ GeV, the cross section $\sigma(\gamma\gamma\to W^+W^-Z)$
is about a factor of two smaller than $\sigma(\gamma\gamma\to t\bar{t})$ \cite{photoncollider},
and the branching ratio of $Z\to b\bar{b}$ is $B(Z\to b\bar{b})=15\%$.
So that this background is smaller than the signal by an order of magnitude.

Another possible background is $\gamma\gamma\to W^+W^-$ in which the hadronization of quarks
forms six jets \cite{JLC}. In $e^+e^-$ collision, the background cross section
$\sigma(e^+e^-\to W^+W^-)$ is much larger than
the signal cross section $\sigma(e^+e^-\to t\bar{t})$ \cite{photoncollider}. However, after taking 
the cut (\ref{ycut}), the signal-to-background ratio can be made greater than 10 \cite{JLC}. At the 
photon collider with $\sqrt{s}=500$ GeV, the signal cross section $\sigma(\gamma\gamma\to t\bar{t})$
is about the same as $\sigma(e^+e^-\to t\bar{t})$, while the background cross section
$\sigma(\gamma\gamma\to W^+W^-)$ is about a factor of 10 larger than $\sigma(e^+e^-\to W^+W^-)$
\cite{photoncollider}. Hence, at the photon collider, the signal-to-background ratio after imposing
the cut (\ref{ycut}) is about 1. So we have to tag at least one $b$-jet to suppress the
background. We know that the $b$-tagging efficiency at LEP and at the Tevatron is around $(50\-- 60)\%$.
We shall take $50\%$ for the $b$-tagging efficiency in the following study.

Our calculation shows that the cut (\ref{ycut}) reduces the signal cross section
$\sigma(\gamma\gamma\to t\bar{t})$ by a factor of $47\%$ in both the SM and
the technicolor models for the parameter range under consideration. 
To be more realistic, we take into account the fact that the detector cannot detect jets
in certain forward and backward zones along the beam line. As a conservative estimate, we take the
polar angle (relative to the beam line) of the undetectable zones to be $\theta<20^\circ$ and 
$\theta>160^\circ$. So we require all the six jets to be in the detectable region 
$20^\circ<\theta<160^\circ$. Practically, $b$-tagging is effective only in the region  
$30^\circ<\theta<150^\circ$. So we further require the tagged $b$-jet and/or $\bar{b}$-jet 
to be in this effective region. There can be two possible cases:
\begin{description}
\item{(a)} Both $b$ and $\bar{b}$ jets are in the $b$-tagging effective region 
$30^\circ<\theta<150^\circ$, while the other four jets are in the detectable region 
$20^\circ<\theta<160^\circ$. Our calculation shows that the probability of satisfying this requirement 
is $58.3\%$. In this case, it is possible to tag both
the $b$ and the $\bar{b}$ jets. Now we only need to tag one of them  
without specifying whether it is $b$ or $\bar{b}$. For each one of them, 
the probability of not being tagged is $50\%$. So the probability of both of them not being tagged is
$25\%$. Therefore our actual $b$-tagging efficiency in this case is $75\%$.
\item{(b)} One tagged $b$ (or $\bar{b}$) jet is in the $b$-tagging effective region 
$30^\circ<\theta<150^\circ$, 
and the untagged $\bar{b}$ (or $b$) jet is in the detectable but not $b$-tagging effective region 
$20^\circ<\theta<30^\circ$. The other four jets are in the detectable region 
$20^\circ<\theta<160^\circ$. Our calculation shows that the probability of satisfying this requirement 
is $10\%$. In this case the $b$-tagging efficiency is $50\%$.
\end{description}
Taking into account both of these two possible cases and the $W$ decay branchiong 
ratio $B(W\to q\bar{q}')=68\%$, our final detection efficiency is
\begin{eqnarray}                            
(68\%)^2\times 47\%\times (58.3\%\times 75\%+10\%\times 50\%)\approx 10\%. 
\label{efficiency}
\end{eqnarray}

In recent years, two of the $e^+e^-$ linear collider projects have
been actively pushed. They are the DESY TeV Energy Superconducting
Linear Accelerator (TESLA) with the designed luminosity of
$3.4\times 10^{34}$ cm$^{-2}$sec$^{-1}$ \cite{TESLA-I} corresponding to
$\int_{yr} {\cal L}dt=340$ fb$^{-1}$, and the KEK
Joint Linear Collider (JLC) with the designed luminosity of
$(8\--9)\times 10^{33}$ cm$^{-2}$sec$^{-1}$ \cite{JLC} corresponding to
$\int_{yr}{\cal L}dt=(80\--90)$ fb$^{-1}$. 
As usual, we shall take an integrated luminosity of 500 fb$^{-1}$ and
the $10\%$ detection efficiency to estimate the numbers of events ($N_{events}$)
of $\gamma\gamma\to t\bar{t}$. In this paper, only statistical errors are taken into account.

 In the following, we calculate the
helicity amplitudes in Eq. (\ref{amplitude}) for $\gamma\gamma\to
t\bar{t}$ in various improved technicolor models. As what we did
in Ref.\cite{ZKYWL}, for avoiding singularities arising from the
very forward or very backward scatterings, we take the rapidity
and transverse momentum cuts
\begin{eqnarray}                        
|y|<2.5,~~~~~~~~p_T>20~{\rm GeV} \label{cuts}
\end{eqnarray}
which will also increase the relative corrections.

As in Ref.\cite{ZKYWL}, we study the PGB contributions to
$\gamma\gamma\to t\bar{t}$ in three technicolor models, namely
Model A: the one-family walking technicolor model by Appelquist
and Terning \cite{AT}, Model B: the top-color-assisted technicolor
model by Hill \cite{TOPCTC}, and Model C: the top-color-assited
multiscale technicolor model by Lane \cite{TOPCMTC}. The PGBs in
these three models are quite different. This is the main reason
why the three models can be distinguished. The formulas for the
PGB-$t\bar{t}$ couplings in the three models and the production
amplitudes are given in Ref.\cite{ZKYWL}.

\null\noindent {\bf Model A}:

 In Model A, color-singlet PGBs are composed of technileptons, which do not
couple to $t\bar{t}$. Thus there is no $s$-channel resonance
contribution in this model. The relevant PGBs are the color-octet
technipions containing the $SU(2)_W$-singlet $\Pi^0_a$ and the
$SU(2)_W$-triplet $\Pi^\alpha_a$. Their masses are in the few
hundred GeV range, and their decay constant is $f_Q\approx 140$
GeV \cite{AT}. The coupling of these PGBs to $t(b)$-quark is
\cite{AT}
\begin{eqnarray}                                 
&&\displaystyle  \frac{\sqrt{2}m_t}{f_Q}\big( i\bar{t}\gamma_5
\frac{\lambda^a}{2}t\Pi^0_a+i\bar{t}\gamma_5
\frac{\lambda^a}{2}t\Pi^3_a +\frac{1}{\sqrt{2}}\bar{t}(1-\gamma_5)
\frac{\lambda^a}{2}b\Pi^+_a\nonumber\\
&&\hspace{1cm}+\frac{1}{\sqrt{2}}\bar{b}(1+\gamma_5)
\frac{\lambda^a}{2}t\Pi^-_a \big). \label{AT-Pitt}
\end{eqnarray}
\begin{table}[h]
\centering\caption{ Technipion corrections to the $\gamma\gamma\to
t\bar{t}$ cross section $\Delta\sigma$, the relative correction
$\Delta\sigma/\sigma_0$, the total cross section
$\sigma=\sigma_0+\Delta\sigma$, and $N_{events}$ for an integrated luminosity
of 500 fb$^{-1}$ taking account of the $10\%$ detection efficiency at a 500 GeV $e^+e^-$ linear
colliderfor with $2\lambda_e P_c=-1$ for various values of
$m_{\Pi_a}$ in Model A (The SM cross section is $\sigma_0=196$
fb).} \label{table-wtc}
\vspace{0.2cm}
\tabcolsep 4pt
\begin{tabular}{ccccc}
\hline\hline
$m_{\Pi_a}$(GeV)& $\Delta \sigma $ (fb) & $\Delta \sigma
/\sigma_0$ ($\%$) & $\sigma $ (fb)& $N_{events}$\\
\hline
250&-31&-15.8&165&8250
\\ 
300&-26&-13.3&170&8500
\\ 
350&-22&-11.2&174&8700
\\ 
400&-19&-9.7&177&8850
\\ 
500&-14&-7.1&182&9100
\\
\hline\hline
\end{tabular}
\end{table}

These color-octet PGBs contribute to $\gamma\gamma\to t\bar{t}$
only through radiative corrections which are small. We list, in TABLE I,
the obtained $\Pi_a$ corrections to the $\gamma\gamma\to t\bar{t}$ cross
section $\Delta\sigma$, the relative correction
$\Delta\sigma/\sigma_0$ ($\sigma_0$ stands for the SM cross
section), the total cross section
$\sigma=\sigma_0+\Delta\sigma$, and $N_{events}$ for an integrated
luminosity of 500 fb$^{-1}$ at a polarized photon collider from
a 500 GeV LC with the polarization $2\lambda_e P_c=-1$ for 
$m_{\Pi_a}$ in the range $250~{\rm GeV}< m_{\Pi_a}< 500~{\rm GeV}$ given in Ref. \cite{AT}.
We see that the corrections $\Delta\sigma$ are negative which are mainly from the
interference between the PGB-amplitude and the SM-amplitude. The
absolute square of the PGB-amplitude is not large in this model.
We see from TABLE I that the total cross sections $\sigma$ are
much larger than those given in Ref.\cite{ZKYWL} due to the
enhancement of the effective photon energy in the high energy
region at the polarized photon collider.
We see that $N_{events}$ for an integrated luminosity of
500 fb$^{-1}$ taking account of the $10\%$ detection efficiency are
$(8\--9)\times 10^3$. The corresponding $95\%$ statistical uncertainties are
about $2\%$.  Comparing with the
relative corrections $\Delta\sigma/\sigma_0\sim -(7\--16)\%$ 
listed in TABLE I, we see that the $\Pi_a$ correction effect can be clearly detected.

\null\noindent {\bf Model B}:

 Model B contains a technicolor sector and a top-color sector. The technicolor sector only
contributes a small portion of the the top quark mass, say
$m_t^\prime=\epsilon m_t$ ($\epsilon\ll 1$), while most of the top
quark mass is contributed from the top-color sector, say
$m_t-m_t^\prime=(1-\epsilon)m_t$. The $b\to s\gamma$ experiment
requires $\epsilon < 0.1$ \cite{Balaji}. In this paper, we take a
typical value $\epsilon=0.08$, i.e., $m_t^\prime=0.08m_t\approx
14$ GeV as an example to do the study.

In the technicolor sector, the coupling of the color-octet
technipion to $t(b)$-quark is similar to Eq. (\ref{AT-Pitt}) but
with $m_t/f_Q$ replaced by $m_t^\prime/f_\Pi$ \cite{ZKYWL} , i.e.,
\begin{eqnarray}                                    
&&\displaystyle  \frac{\sqrt{2}m_t^\prime}{f_\Pi}\bigg(
i\bar{t}\gamma_5 \frac{\lambda^a}{2}t\Pi^0_a+i\bar{t}\gamma_5
\frac{\lambda^a}{2}t\Pi^3_a +\frac{1}{\sqrt{2}}\bar{t}(1-\gamma_5)
\frac{\lambda^a}{2}b\Pi^+_a\nonumber\\
&&\hspace{0.8cm}+\frac{1}{\sqrt{2}}\bar{b}(1+\gamma_5)
\frac{\lambda^a}{2}t\Pi^-_a \bigg). \label{TOPCTC-Piatt}
\end{eqnarray}
In addition to the color-octet technipions, there are also
color-singlet technipions $\Pi^0$ and $\Pi^3$, composed of
techniquarks, with the decay constant $F_\Pi\approx 120$ GeV, and
masses in the few hundred GeV range. The coupling of these
color-singlet PGBs to $t(b)$-quarks is \cite{ZKYWL}
\begin{eqnarray}                                    
&&\displaystyle  \frac{c_tm'_t}{\sqrt{2}f_{\Pi}}\bigg(
i\bar{t}\gamma_5 t\Pi^0+i\bar{t}\gamma_5 t\Pi^3
+\frac{1}{\sqrt{2}}\bar{t}(1-\gamma_5) b\Pi^+\nonumber\\
&&\hspace{0.8cm} +\frac{1}{\sqrt{2}}\bar{b}(1+\gamma_5) t\Pi^-
\bigg), ~~~~~~c_t=1/\sqrt{6}.\label{TOPCTC-Pi0tt}
\end{eqnarray}

In the top-color sector, there are color-singlet top-pions
$\Pi_t^0$ and $\Pi_t^\pm$ with the decay constant
$F_{\Pi_t}\approx 50$ GeV. The coupling of top-pions to
$t(b)$-quark is \cite{ZKYWL}
\begin{eqnarray}                                  
&&\frac{m_t-m_t'}{\sqrt{2}f_{\Pi_t}}\bigg( i\bar{t}\gamma_5
t\Pi_t^{0} +\frac{1}{\sqrt{2}}\bar{t}(1-\gamma_5)
b\Pi_t^{+}\nonumber\\
&&\hspace{0.8cm}+\frac{1}{\sqrt{2}}\bar{b}(1+\gamma_5) t\Pi_t^{-}
\bigg). \label{TOPCTC-Pittt}
\end{eqnarray}

Recently, it has been pointed out that the LEP precision data of
$R_b$ gives important constraint on the top-pion mass \cite{YKWL}.
With $\epsilon=0.08$, the $2\sigma$ bound from $R_b$ on the top-pion mass in
this model is roughly $300~{\rm GeV}\le m_{\Pi_t}\le 900$ GeV \cite{YKWL}.
The naturalness of the model favors lower values of $m_{\Pi_t}$. So we take
$300~{\rm GeV}\le m_{\Pi_t}\le 500~{\rm GeV}$ in this study. In
Ref.\cite{ZKYWL}, $m_{\Pi_t}$ was taken to be in the range of
$180\--300$ GeV according to the original paper,
Ref.\cite{TOPCTC}. Such a range is below the recent lower bound.
We shall see that the updated heavier $\Pi_t$ will make the
situation quite different at the polarized photon collider from
the 500 GeV LC.

The color-singlet PGBs $\Pi^0$, $\Pi^3$ and $\Pi_t^0$ couple to
the initial state photons through triangle fermion loops
(techniquark loops and top quark loops), so they can contribute
$s$-channel resonances in $\gamma\gamma\to t\bar{t}$. The triangle
fermion loops are enhanced by the anomaly, so that the $s$-channel
resonance contributions can be of the order of tree level
contributions. These $s$-channel resonance contributions are
dominant in Model B and Model C. If the PGB mass is greater than
$2m_t$, it can decay into $t\bar{t}$. The decay rate is determined
by the PGB-$t-\bar{t}$ coupling strengths given in Eqs.
(\ref{TOPCTC-Pi0tt}) and (\ref{TOPCTC-Pittt}). Since
$m_t-m_t^\prime/f_{\Pi_t}$ in Eq. (\ref{TOPCTC-Pittt}) is large,
the width of $\Pi_t^0$ will be large if its mass is greater than
$2m_t$. In this case, the $s$-channel resonance effect from
$\Pi^0_t$ is not so significant. On the other hand,
$m_t^\prime/f_\pi$ in Eq. (\ref{TOPCTC-Pi0tt}) is small, so that
the resonance effects from $\Pi^0$ and $\Pi^3$ are significant
even their masses are greater than $2m_t$. The width of $\Pi^3$ is
very small which is hard to detect experimentally \cite{ZKYWL}. So
we concentrate on examining the $s$-channel resonance effect of
$\Pi^0$, and simply take a typical mass of $m_{\Pi^3}=300$ GeV for
$\Pi^3$.
\begin{table}[h]
\centering\caption{Technipion and top-pion corrections to the
$\gamma\gamma\to t\bar{t}$ cross section $\Delta\sigma$, the
relative correction $\Delta\sigma/\sigma_0$, the total cross
section $\sigma=\sigma_0+\Delta\sigma$, and $N_{events}$ for an
integrated luminosity of 500 fb$^{-1}$ taking account of the $10\%$
detection efficiency at a 500 GeV $e^+e^-$
linear colliderfor with $2\lambda_e P_c=-1$ for $m_t^\prime$=14
GeV, $m_{\Pi^3}=300$ GeV, and various values of $m_{\Pi_t^0}$ and
$m_{\Pi^0}$ in Model B (The SM cross section is $\sigma_0=196$
fb).} \label{table-topctc}
\begin{tabular}{ccccc}
$m_{\Pi_t}=300$GeV
\\ \hline\hline
$m_{\Pi^0}$& $\Delta \sigma $ (fb) & $\Delta \sigma /\sigma_0$
($\%$) & $\sigma $ (fb)&$N_{events}$
\\ \hline
300&-86.2& -44.0 & 110&5500\\
400&-47.2& -24.1& 149&7450\\
500&-87.4& -44.6 & 109&5450\\
\hline\hline\\
$m_{\Pi_t}=400$GeV
\\ \hline\hline
$m_{\Pi^0}$& $\Delta \sigma $ (fb) & $\Delta \sigma /\sigma_0$
($\%$) & $\sigma $ (fb)&$N_{events}$
\\ \hline
300&112&57.1 &308&15400\\ 
400&158.3&80.8 &354&17700\\ 
500&115&58.7&311&15550\\
\hline\hline
\end{tabular}
\end{table}

The obtained $\Delta\sigma$, $~\Delta\sigma/\sigma_0$,
$~\sigma$, and $N_{events}$ for an intefrated lumionosity of 500 fb$^{-1}$
taking account of the $10\%$ detection efficiency  for
$m_{\Pi_t^0}$=300 and 400 GeV, and $m_{\Pi^0}$=300,
~400, and ~500 GeV in this model are listed in TABLE II. We see
that, in the case of $m_{\Pi_t^0}$=300 GeV, $\Delta\sigma$ is more
negative than in Model A because there is $s$-channel $\Pi_t^0$
and $\Pi^0$ resonance contributions in addition to the radiative
corrections, and the $s$-channel resonance contributions are
mainly from the interferences between the PGB-amplitudes and the
SM-amplitude. In the case of $m_{\Pi_t^0}$=400 GeV, $\Delta\sigma$
becomes positive. This is because that the absolute squares of the
$\Pi_t$ and $\Pi^0$ resonance amplitudes are large in this case
due to the enhancement of the photon spectral luminosity in the
region around 400 GeV ($80\%$ 0f the $e^+e^-$ energy) in the case
of $2\lambda_eP_c=-1$ at the polarized photon collider
\cite{Ginzburg}.
\begin{figure}[h]
\includegraphics[width=8truecm,clip=true]{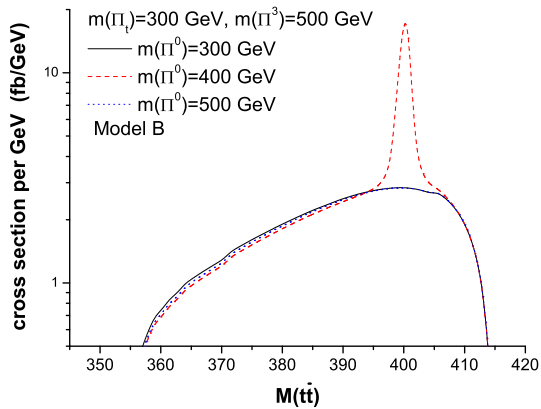}
\null\vspace{-0.8cm}\caption{$t\bar{t}$ invariant mass distribution in Model B for
$m_{\Pi_t}$=300 GeV and $m_\Pi$=300, 400, and 500 GeV.}
\end{figure}

To have an insight of the detailed situation, we plot the
$t\bar{t}$ invariant mass distributions in Fig. 1 and Fig. 2 for
the six sets of parameters in TABLE II. Fig. 1 shows the invariant
mass distributions for $m_{\Pi_t}$=300 GeV, $m_\Pi$=300, 400, and 500 GeV.
We see that all curves are
enhanced in the region around 400 GeV which is just the effect of
the photon spectral luminosity. In the $m_{\Pi^0}=400$ GeV
distribution, we see a clear peak at 400 GeV.
In the $m_{\Pi^0}=300$ and 500 GeV distributions no clear peaks can be
seen. This is because that the probability of the center-of-mass
energy of the two colliding photons being 300 or 500 GeV is very
small as can be seen from the photon spectral luminosity
\cite{Ginzburg}. We cannot see the top-pion resonance since the
width of top-pion is much larger. Fig. 2 shows the invariant mass
distributions for $m_{\Pi_t}$=400 GeV. The behaviors are similar
but the enhancement around 400 GeV is stronger due to the
$s$-channel contribution of the $m_{\Pi_t}=400$ GeV top-pion.
\begin{figure}[h]
\includegraphics[width=8truecm,clip=true]{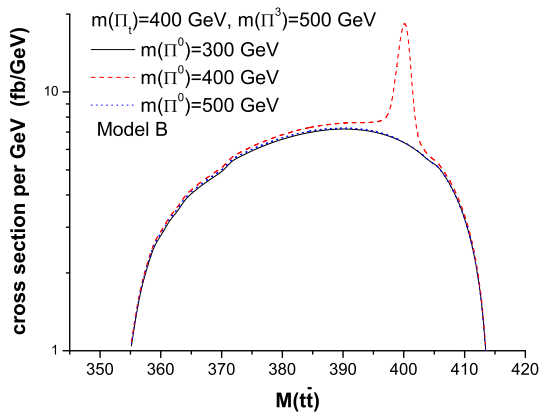}
\null\vspace{-0.8cm}\caption{$t\bar{t}$ invariant mass distribution in Model B for
$m_{\Pi_t}$=400 GeV and $m_\Pi$=300, 400, and 500 GeV.}
\end{figure}

In the case of $m_{\Pi_t}$=300 GeV,  
$N_{events}$ is $(5.4\--7.5)\times 10^3$ (cf. TABLE II). The $95\%$ C.L. statistical
uncertainties are thus $(2\--3)\%$. 
Comparing with the relative corrections
$\Delta\sigma/\sigma_0=(24\--45)\%$ in TABLE II, we see that these
correction effects can be very clearly detected. For distinguishing Model B
from Model A, we see that the relative
difference between the total cross sections in Model A and Model B
is $[\sigma(A)-\sigma(B)]/\sigma(A)=(15\--40)\%$, so that these
two models can be very clearly distinguished.

In the case of $m_{\Pi_t}$=400 GeV, $N_{events}$ is 
$(1.5\--1.7)\times 10^4$. 
The corresponding $95\%$ C.L. statistical uncertainties are $(1.5\--1.6)\%$.  
Now the relative corrections are
$\Delta\sigma/\sigma_0=(57\--81)\%$ in TABLE II, so that these
correction effects can be very clearly detected. The relative
difference between the total cross sections in Model A and Model B
is now $[\sigma(A)-\sigma(B)]/\sigma(A)=-(71\--100)\%$, thus these
two models can be very clearly distinguished.

\null\noindent {\bf Model C}:

 Model C is similar to Model B, but
the decay constant of the color-singlet technipions is $f_\Pi=40$
GeV rather than $f_\Pi\approx 120$ GeV \cite{TOPCMTC}. Moreover, the
coupling constant $c_t$ in Eq. (\ref{TOPCTC-Pi0tt}) and the
$\Pi^0$-$\gamma$-$\gamma$ coupling are also different from those in
Model B \cite{TOPCMTC,ZKYWL}. The smallness of $f_\Pi$ makes the
coupling constants $m_t^\prime/f_\Pi$ in Eqs. (\ref{TOPCTC-Piatt})
and (\ref{TOPCTC-Pi0tt}) larger than those in Model B by a factor
of 3. These changes enhance the the $s$-channel
$\Pi^0$ resonance effect in Model C significantly. Furthermore, the $2\sigma$
$R_b$ bound on the top-pion mass in this model is roughly
$250~{\rm GeV}\le m_{\Pi_t}\le 560~{\rm GeV}$ \cite{YKWL}. To compare with
Model B, we also take the range $300~{\rm GeV}\le m_{\Pi_t}\le 500~{\rm GeV}$
in this study.

\null\vspace{-0.6cm}
\begin{table}[h]
\centering\caption{Technipion and top-pion corrections to the
$\gamma\gamma\to t\bar{t}$ cross section $\Delta\sigma$, the
relative correction $\Delta\sigma/\sigma_0$, the total cross
section $\sigma=\sigma_0+\Delta\sigma$, and $N_{events}$ for
an integrated luminosity of 500 fb$^{-1}$ taking account of the $10\%$
detection efficiency at a 500 GeV $e^+e^-$
linear colliderfor with $2\lambda_e P_c=-1$ for $m_t^\prime$=14
GeV and various values of $m_{\Pi_t}$ and $m_{\Pi^0}$ in Model C
with $m_{\Pi^3}=300$ GeV (The SM cross section is $\sigma_0=196$
fb).} \label{table-topcmtc}
\tabcolsep 2pt
\begin{tabular}{ccccc}
$m_{\Pi_t}=300$GeV
\\ \hline\hline
$m_{\Pi^0}$& $\Delta \sigma $ (fb) & $\Delta \sigma /\sigma_0$
($\%$) & $\sigma $ (fb)&$N_{events}$
\\ \hline
300&-22.1&-11.3  & 174&8700
\\ 
400&469.2&239.4 & 665&33250
\\ 
500&-92.3&-47.1  & 104&5200
\\\hline\hline\\
$m_{\Pi_t}=400$GeV
\\ \hline\hline
$m_{\Pi^0}$& $\Delta \sigma $ (fb) & $\Delta \sigma /\sigma_0$
(fb) & $\sigma $ (fb)&$N_{events}$
\\ \hline
300&79.1&40.4 &  275&13750
\\ 
400&1647.3&840.5 &  1843&92150
\\ 
500&125.7&64.1 &  322&16100
\\\hline\hline
\end{tabular}
\end{table}

As in the case of Model B, we examine the cases of $m_{\Pi_t}=300$
and 400 GeV, and $m_{\Pi^0}=300$, 400, and 500 GeV, with a typical
value of the $\Pi^3$ mass, $m_{\Pi^3}=$ 300 GeV. The obtained
values of $\Delta\sigma$, $\Delta\sigma/\sigma_0$, and $\sigma$
are listed in TABLE III. We see that $\Delta\sigma$ is negative
only in the case of $m_{\Pi_t}=m_{\Pi^0}=300$ GeV. In all other
cases, $\Delta\sigma$ is positive because the absolute square of
the $\Pi^0$ resonance amplitude is large due to the largeness of
the coupling constant $m_t^\prime/f_\Pi$. This effect is very
significant in the case of $m_{\Pi_t}=m_{\Pi^0}=400$ GeV.
\begin{figure}[t]
\includegraphics[width=8truecm,clip=true]{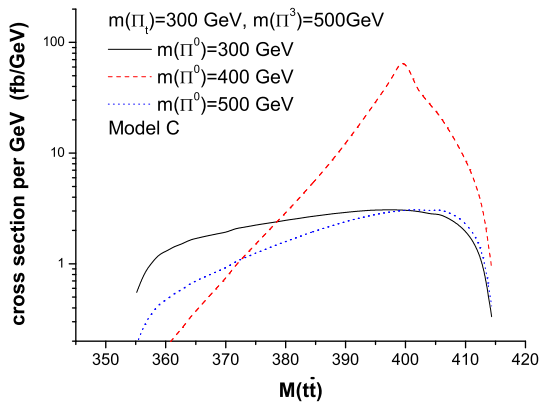}
\null\vspace{-0.8cm}\caption{$t\bar{t}$ invariant mass distribution in Model C for
$m_{\Pi_t}$=300 GeV and $m_\Pi$=300, 400, and 500 GeV.}
\end{figure}

In Figs. 3 and 4, we plot the $t\bar{t}$ invariant mass
distributions in Model C for $m_{\Pi_t}=$ 300~ and~ 400 GeV and
$m_{\Pi^0}$= 300, 400, and 500 GeV. Again, we see the clear
$\Pi^0$ resonance peak at 400 GeV, but the width of the resonance
is much larger than that in Figs. 1 and 2 due to the largeness of
the $\Pi^0\to {\rm gluons}$ and $\Pi^0\to t\bar{t}$ widths.
\begin{figure}[h]
\includegraphics[width=8truecm,clip=true]{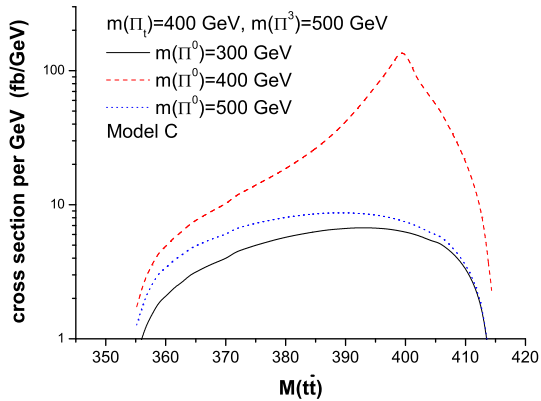}
\null\vspace{-0.8cm}\caption{$t\bar{t}$ invariant mass distribution in Model C for
$m_{\Pi_t}$=400 GeV and $m_\Pi$=300, 400, and 500 GeV.}
\end{figure}

In the case of $m_{\Pi_t}$=300 GeV, $N_{events}$ is $(5.2\--33)\times 10^3$.
The corresponding $95\%$ C.L. statistical
uncertainties are then $(1\--3)\%$. 
Compared with the large relative
corrections $\Delta\sigma/\sigma_0$ listed in TABLE III, we see
that the PGB effects in Model C can be clearly detected. Comparing
the relative corrections $\Delta\sigma/\sigma_0$ listed in TABLE
III  and TABLE II, we see that the difference between Model C and
Model B is significant for $m_{\Pi^0}$=300 and 400 GeV. For
$m_{\Pi^0}$=500 GeV, the relative difference is
$[\sigma(B)=\sigma(C)]/\sigma(B)=[109-104]/109=5\%$ which is also
beyond the statistical uncertainties. Thus model C
and Model B can be clearly distinguished from the production cross
sections.
\begin{figure}[h]
\includegraphics[width=8truecm,clip=true]{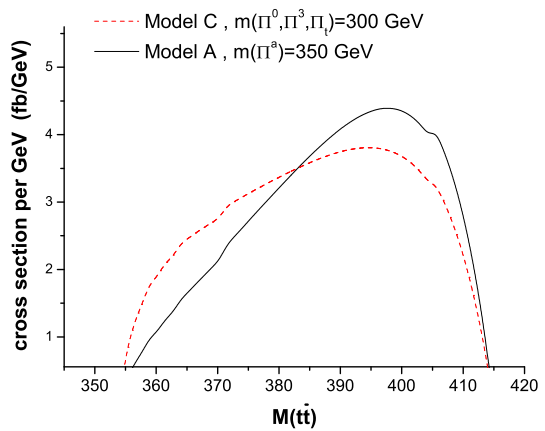}
\null\vspace{-0.8cm}\caption{Comparison of the $t\bar{t}$ invariant mass distributions
in Model C for $m_{\Pi_t}=m_\Pi=m_{\Pi_a}$=300 GeV and Model A for
$m_{\Pi_a}=350$ GeV.}
\label{ACS}
\end{figure}

We see from TABLE III and TABLE I that, for most values of the PGB masses, Model C
can be distinguished from Model A. 
The only case which needs to be studied more carefully is
distinguishing the cases of Model C with $m_{\Pi_t}=m_{\Pi^0}$=300
GeV from Model A with $m_{\Pi_a}$=350$\--$400 GeV. From TABLE III and TABLE
I, we see that the total cross sections for these two cases are
almost the same, so that they cannot be distinguished by merely measuring
the total cross sections. Since the numbers of events are around 8700$\--$8850, it is posiible
to measure the $t\bar{t}$ invariant mass ($M_{t\bar{t}}$) distribution. In Fig. 5 we plot the 
$M_{t\bar{t}}$ distributions for the two cases.
\begin{figure}[h]
\includegraphics[width=8truecm,clip=true]{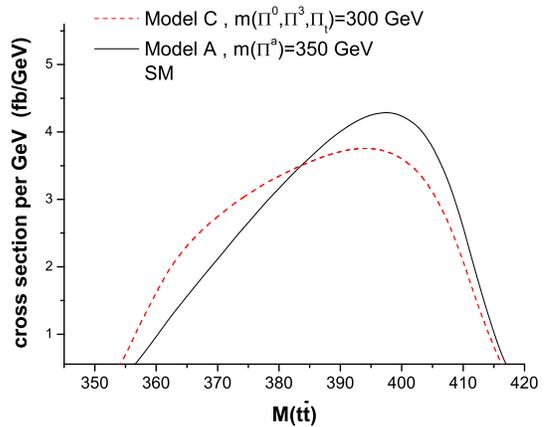}
\null\vspace{-0.8cm}\caption{Comparison of the smeared $t\bar{t}$  invariant mass distributions
in Model C for $m_{\Pi_t}=m_\Pi=m_{\Pi_a}$=300 GeV and Model A for
$m_{\Pi_a}=350$ GeV according to the resolution ${\Delta E}/E=0.33/\sqrt{E}$.}
\label{ACsmear}
\end{figure}

To make it more realistic, we should furhter take into account the effect of energy resolution.
According to Ref. \cite{TESLA-IV}, the energy resolution is $\Delta E/E=0.33/\sqrt{E}$. We then
smear the calculated $M_{t\bar{t}}$ by taking convolutions with this resolution.
The obtained smeared $M_{t\bar{t}}$ distrobutions for the above two cases are shown 
in Fig. 6.
We see that they are different, especially in the vicinities of 365 GeV and 400 GeV where
the differences are of the level of $(10\--40)\%$.
According to the energy resolution, the resolution for measuring the $M_{t\bar{t}}$ distribution 
around 400 GeV is $\Delta M_{t\bar{t}}=0.33\times \sqrt{400}~{\rm GeV}=6.6$ GeV. So that the above 
two cases in model A and Model C can be distinguished by separately measuring the numbers of events
in the regions $355~{\rm GeV}\le M_{t\bar{t}}\le 375~{\rm GeV}$ and 
$390~{\rm GeV}\le M_{t\bar{t}}\le 410~{\rm GeV}$.

In the case of $m_{\Pi_t}$=400 GeV, we can see that Model C is
very different from the SM, and all the three models can be distinguished.

 In summary, we have studied the possibility of
testing and distinguishing three improved technicolor models
(Models A, B, and C) via $t\bar{t}$ productions at a polarized
photon collider with $2\lambda_eP_c=-1$ from a 500 GeV $e^+e^-$ linear collider with an
integrated luminosity of 500 fb$^{-1}$. The signal contains six jets
from $t\to W^+b$ and $W\to q\bar{q}'$. Backgronds can be suppressed
by taking the cut (\ref{ycut}) and tagging a $b$ quark jet. Considering the possible detection 
ability of the detector and the usual $b$-tagging efficiency, the detection
efficiency is $10\%$.
We see that, considering only the statistical error, the three improved technicolor models can
all be well tested and can be distinguished from each
other by measuring the $\gamma\gamma\to t\bar{t}$ production cross
section and the $t\bar{t}$ invariant mass distribution.

\null\noindent {\bf Acknowledgement}: We would like to thank C.-P.
Yuan for discussions.  This work is supported by the National
Natural Science Foundation of China under the grant number
90103008.


\begin{thebibliography}{000}

\bibitem{PDG}
S. Eidelman, {\it al}., (Particle Data Group), Phys. Lett. {\bf B 592}, 1 (2004).
\bibitem{AT}
T. Appelquist and J. Terning, Phys. Lett. {\bf B 315}, 139 (1993).
\bibitem{TOPCTC}
C.T. Hill, Phys. Lett. {\bf B 345}, 483 (1995); K. Lane and E.
Eichten, {\it ibit}. {\bf 352}, 382 (1995); G. Buchalla, G.
Burdan, C.T. Hill, and D. Kominis, Phys. Rev. D {\bf 53}, 5185
(1996).
\bibitem{TOPCMTC}
K. Lane, Phys. Lett. {\bf B 357}, 624 (1995).
\bibitem{seesaw}
B. A. Dobrescu and C. T. Hill, Phys. Rev. Lett. {\bf 81}, 2634
(1998); R. S. Chivukula, B. A. Dobrescu, H. Georgi, and C. T.
Hill, Phys. Rev. D {\bf 59}, 075003 (1999); H.-J. He, C. T. Hill,
and T. Tait, Phys. Rev. D {\bf 65}, 055006 (2002);
\bibitem{littleHiggs}
N. Arkani-Hammed, A.G. Cohen, and H. Georgi, Phys. Lett. {\bf B
513} (2001) 232; N. Arkani-Hammed, A.G. Chen, E. Katz, A.E.
Nelson, T Gregoire, and J.G. Wacker, JHEP {\bf 0208} (2002) 021;
J.G. Wacker, hep-ph/0208235; N. Arkani-Hammed, A.G. Cohen, E.
Katz, and A.E. Nelson, JHEP {\bf 0207} (2002) 034; T. Gregoire and
J.G. Wacker, hep-ph/0207164; I. Low, W. Skiba, and D. Smith,
hep-ph/0207243.
\bibitem{ZKYWL}
H.-Y. Zhou, Y.-P. Kuang, C.-X. Yue, H. Wang, and G.-R. Lu, Phys.
Rev. D {\bf 57}, 4205 (1998).
\bibitem{YKWL}
C.-X. Yue, Y.-P. Kuang, X.-L. Wang, and W.B. Li, phys. Rev. D {\bf
62}, 055005 (2000).
\bibitem{Ginzburg}
I.F.Ginzburg,G.L. Kotkin, S.L. Panfil, V.G. Serbo, and V.I.
Telnov, Nucl. Instr. and Methods in Phys. Res. {\bf 219}, 5
(1984); B. Badelek {\it et al}., Part VI of TESLA Technical Design
Report, DESY 2001-011.
\bibitem{luminosity}
O.J.P. Eb\'oli {\it et al}., Phys. Rev. D {\bf 47}, 1889 (1993);
K. Cheung,, {\it ibid}. {\bf 47}, 3750 (1993).
\bibitem{BGMS}
V.M. Budnev, I.F. Ginzburg, G.V. Meledin, and V.G. Serbo, Phys.
Rep. {\bf 15 C}, 181 (1975).
\bibitem{JLC}
{\it Particle Physics Experiments at JLC} (JLC Technical Design
Report), KEK Report 2001-11.
\bibitem{photoncollider}
M. Baillargeon {\it et al}., in {\it $e^+e^-$ Collisions at TeV Energies:
The Physics Potential, Part D}, edited by P. Zerwas, DESY 96-123D (1995);
see also Part VI of {\it TESLA Technical Design Report}, DESY 2001-011.
\bibitem{TESLA-I}
Part I of {\it TESLA Technical Design Report}, DESY 2001-011.
\bibitem{Balaji}
B. Balaji, Phys. Rev. D {\bf 53}, 1699 (1996).
\bibitem{TESLA-IV}
Part IV of {\it TESLA Technical Design Report}, DESY 2001-011.
\end{thebibliography}
\end{document}